\documentclass[10pt,manuscript,nonacm]{acmart}
\usepackage[T1]{fontenc}
\usepackage{lmodern}
\usepackage[utf8]{inputenc}
\usepackage{hyperref}
\usepackage{graphicx} 
\usepackage[english]{babel}
\usepackage{setspace}
\usepackage{booktabs}
\usepackage{subcaption}

\newcommand{\maxblockheight}{840,682}
\newcommand{\maxnbaddresses}{1,283,997,050}
\newcommand{\maxnbaddressesnosent}{42,964,903}

\begin{document}
\title{\textbf{Bitcoin Burn Addresses: Unveiling the Permanent Losses and Their Underlying Causes}}

\author{Mohamed El Khatib}
\affiliation{%
    \institution{Centre Inria d'Université Côte d'Azur}
    \city{Sophia Antipolis}
    \country{France}}
\author{Arnaud Legout}
\authornote{Corresponding author. Contact email: arnaud.legout@inria.fr}
\orcid{0000-0003-4706-424X}
\affiliation{%
    \institution{Centre Inria d'Université Côte d'Azur}
    \city{Sophia Antipolis}
    \country{France}}

\begin{abstract}
    Bitcoin burn addresses are addresses where bitcoins can be sent but never
    retrieved, resulting in the permanent loss of those coins. Given Bitcoin's
    fixed supply of 21 million coins, understanding the usage and the amount of
    bitcoins lost in burn addresses is crucial for evaluating their economic
    impact. However, identifying burn addresses is challenging due to the lack
    of standardized format or convention.

    In this paper, we propose a novel methodology for the automatic detection of
    burn addresses using a multi-layer perceptron model trained on a manually
    classified dataset of 196,088 regular addresses and 2,082 burn addresses.
    Our model identified 7,905 true burn addresses from a pool of
    1,283,997,050 addresses with only 1,767 false positive. We determined
    that 3,197.61 bitcoins have been permanently lost, representing only 0.016\%
    of the total supply, yet 295 million USD on November 2024. More than 99\%
    of the lost bitcoins are concentrated in just three addresses. This skewness
    highlights diverse uses of burn addresses, including token creation via
    proof-of-burn, storage of plain text messages, or storage of images using
    the OLGA Stamps protocol.

\end{abstract}

\maketitle
\sloppy

\section{Introduction}
Since the first transaction in January 2009, the usages of
Bitcoin~\cite{nakamoto2009bitcoin} have evolved. One intriguing use is the
creation of burn addresses. A burn address is an address to which bitcoins can
be sent, but from which they can never be retrieved. Therefore, two questions
arise. i) How much money have been definitely lost in burn addresses.
Considering the maximum supply of bitcoins is 21 million, evaluating the amount of
money lost in burn addresses is important. ii) What are burn addresses used for?
Exploring the usage of burn addresses can inform on uses or misuses of the
blockchain.

The main challenge with burn addresses is that they are hard to identify. There
is no standardized format or admitted convention to create a burn address.
However, burn addresses must be created in a way that makes them easily
identifiable by a human, typically, they must look handmade.  Since the address
is derived from a private key using elliptic curve
cryptography~\cite{book-mastering-btc}, the probability of having a private key
that corresponds to an address that appears handmade, e.g.,
\texttt{1CounterpartyXXXXXXXXXXXXXXXUWLpVr}, is extremely low. Therefore, users
can be confident that there is no associated private key for this address,
meaning there is no way to retrieve the money sent to it.

Burn addresses are known to be used for various purposes, such as the creation
of new cryptocurrencies using proof-of-burn~\cite{Karantias2020} or the
permanent recording of a simple message in an address. Because burn addresses are difficult to identify automatically,
there is no public list of these addresses and no study has been conducted to
explore the total amount of bitcoins lost in burn addresses and their usages.

In this paper, we make three contributions.
\begin{enumerate}
    \item  We propose a methodology to automatically detect burn addresses. We manually
          classified 208,656 addresses suspected to be burn addresses because they have a
          low Shannon entropy. Among them, we identified 2,082 burn addresses and 196,088
          regular addresses. We used this ground truth to train a multi-layer perceptron
          (MLP) model that we iteratively reinforced to detect new burn addresses. The
          model identified 5,823 additional true burn addresses that we all manually
          validated. Therefore, we built the largest to date list of burn addresses with a
          total of 7,905 addresses and provide a machine learning model to automatically
          detect new burn addresses. We make the ground truth initial training set, the
          list of all burn addresses manually scrutinized, and the
          model publicly available to the community (see Annex~\ref{sec:appendix_dataset}).

    \item  We explore the amount of money lost in burn addresses. We found 3,197.61
          bitcoins permanently lost in burn addresses, that is 0.016\% of the total
          supply, yet representing  295 million USD on November 2024. However, the distribution
          of the bitcoins among burn addresses is highly skewed. The top 3 burn addresses
          represent more than 99\% of the total amount of bitcoins permanently lost.

    \item  We explore the characteristics and usages of burn addresses. Base58 burn addresses are mostly used to store short to long
          messages in plain text, and Bech32 addresses are used to store images
          or animated GIFs mainly using the OLGA Stamps
          protocol~\cite{stamps-olga}. The addresses that received the largest
          amount of bitcoins are used for a proof-of-burn protocol to create new
          tokens such as XCP for the Counterparty project~\cite{counterparty} or
          to operate a protocol such as Blockstack~\cite{blockstack2016}.
\end{enumerate}

The notion of burned bitcoins is also used in the context of
non-standard transactions~\cite{bistarelli2019analysis} and arbitrary messages
stored in the
blockchain~\cite{Bartoletti2019, matzutt2018quantitative, scheid2023opening, sward2018data}.
However, in all these works the authors specifically focus on the structure of
transactions rather than the structure of addresses. Consequently, our work is
complementary to these works and shed a new light on the usage of burn
addresses. We discuss the related work in Section~\ref{sec:related_work}.

In the following, we provide a detailed description of the methodology
implemented to identify burn addresses in Section~\ref{sec:methodology} and our
results in Section~\ref{sec:results}. More specifically, we present the
performance of the model to identify burn addresses in
Section~\ref{sec:results_model_predictions},  the amount of money lost in burn
addresses in Section~\ref{sec:results_amount_of_bitcoin_lost_in_burn_addresses},
and the characteristics and usages of burn addresses in
Section~\ref{sec:results_burn_address_usages}. We discuss the related work in
Section~\ref{sec:related_work}, and conclude in Section~\ref{sec:conclusion}.

\section{Methodology}
\label{sec:methodology}

Our goal is to design a method to automatically detect burn addresses. Because,
a burn address must be clearly recognizable as forged, we consider a human
identifying an address as burn to be the ground truth\footnote{It is possible to
    have a legitimate address that could be identified as burn because, by chance,
    we can observe a structure that looks hand made. However, as this probability is
    extremely low, we consider this event negligible.}. We can then use this ground
truth to train a machine learning model to automatically detect burn addresses.
Then, we iterate by: i) applying the model on all Bitcoin addresses, ii)
manually checking the addresses identified as burn, iii) adding the new burn
addresses to the training set, iv) retraining the model, and v) repeating the
process until convergence.

All results are derived from the full Bitcoin blockchain up to block height
    {\maxblockheight} mined on April 24, 2024. In the following, we provide a detailed description of the
methodology implemented to identify Bitcoin burn addresses.

\subsection{Background}
\label{subsec:background}
In this section, we introduce the concepts of address encoding, burn addresses,
and vanity addresses. These notions are essential to understand the remainder of
the paper.

\subsubsection{Address encoding}
\label{subsubsec:address_encoding}
Bitcoin addresses can be encoded with two different alphabets: Base58 (using a
58 characters alphabet) and the more recent Bech32~\cite{book-song,BIP173}
(using a 32 characters alphabet). One significant difference between the two
alphabets is that Bech32 is only using lower case characters. Therefore, there
is no possibility to use the CamelCase convention~\cite{camelCase-wikipedia} to write messages in a Bech32
address, a convention largely used for addresses encoded in Base58 as we will
see in the following.

\subsubsection{Burn addresses}
\label{subsubsec:burn_addresses}
To understand what is a Bitcoin burn address, we first need to understand how is
working a Bitcoin address. Anyone can create a new Bitcoin address and there is
no limit on the number of created addresses. The creation process is usually
automatically managed by a wallet software, e.g., Bitcoin
core~\cite{bitcoin-core}, but it can also be done manually. Bitcoin addresses
are based on asymmetric cryptography using elliptic curve cryptography. When a
Bitcoin address needs to be created, a private key is generated randomly. The
public key is then derived from the private key using elliptic curve
cryptography. This is an asymmetric operation as deriving the private key from
the public key is computationally hard. The public key is then hashed using
HASH160 and encoded with a checksum using Base58 or Bech32 encoding  depending
on the address version~\cite{book-mastering-btc}. Once we have a Bitcoin
address, we can send bitcoins to this address. To spend the bitcoins sent to an
address, we need to have the private key that corresponds to the address.
Therefore, without a private key associated to the Bitcoin address, the bitcoins
sent to the address are definitively lost.

A Bitcoin burn address must be provably unspendable, meaning it can be easily
verified that no private key corresponds to it, making the coins unspendable.
Although the term "provably unspendable" is widely used in the community, it is
technically an overstatement since proof of unspendability is not possible for
Bitcoin addresses---only an extremely low probability of spendability can be
asserted. Because generating a private key that corresponds to a given
address is computationally hard (equivalent to reverting elliptic curve
cryptography), burn addresses leverage this property by being noticeably
different from regular addresses. Therefore, if the address looks hand made, we
can be confident that there is no associated private key, so the address is
effectively a burn address. Typically, burn addresses must contain a structure
that looks hand made and cannot be due to chance, e.g.,
\texttt{1CounterpartyXXXXXXXXXXXXXXXUWLpVr} or
\texttt{1KToEddieAndELLieLoveYouDadXWz9SPM}.

This definition of burn addresses as provably unspendable is in the context of
proof-of-burn~\cite{stewart-proof-of-burn, Karantias2020}. However, as we
will discuss in Section~\ref{sec:results_burn_address_usages}, burn addresses
are also used for other purposes. The most notable uses include storing a
message or an image. In this case, the address is indeed unspendable, because
its structure is determined by the data to be stored, not by a private key.
However, the reason of its creation is unrelated to proof-of-burn and it is
not intended to receive bitcoins.

\subsubsection{Vanity addresses}
\label{subsubsec:vanity_addresses}
Vanity addresses~\cite{book-mastering-btc} contain a specific
pattern, so they might look like burn addresses, but they
correspond to valid Bitcoin addresses. In fact, vanity addresses are forged using
brute force to find the associated private key. Therefore, the larger the number
of characters in the pattern, the longer it takes to forge the address.
Assuming we need $s$ seconds to test an address, we need $s \times 58^n$ seconds
to forge an address with $n$ characters in Base58 and $s \times 32^n$
seconds in Bech32. Considering $s=1{\mu}s$, it takes 792,321 years to forge an address with 11
characters in Base58 and 1,169,884 years with 13 characters in Bech32. In the following, we consider addresses with 11 or
more forged characters to be non-vanity for Base58 and addresses with
13 or more forged characters to be non-vanity for Bech32.

\subsection{Initial training set construction}
\label{subsec:initial_training_set_construction}

To extract all Bitcoin addresses, we use the Bitcoin core client~\cite{bitcoin-core} to download
the full blockchain and pycoin~\cite{pycoin} to deserialize all transactions and extract
all addresses.

Up to block height {\maxblockheight} mined on April 24, 2024, there are
    {\maxnbaddresses} Bitcoin addresses. However, by definition, burn addresses
cannot send bitcoins, therefore we can reduce the set of candidate burn
addresses by keeping only addresses that never sent bitcoins, after this
filtering, it remains {\maxnbaddressesnosent} addresses. As the density of burn
addresses is low, we cannot randomly select addresses to create the initial
training set. Considering that burn addresses do not look random, such addresses
should have a lower Shannon entropy \cite{shannon-entropy} than regular
addresses. Thus, we use Shannon entropy to filter out many regular addresses.

We computed the Shannon entropy for all Bitcoin addresses and found that the
majority have an entropy between 4 and 5 (the histogram of the Shannon entropy
for all burn addresses is available in Fig.~\ref{fig:entropy_hist} in
Annex~\ref{sec:appendix_histogram_shannon_entropy}). Consequently, we decided to
classify all addresses with an entropy strictly lower than 4 as the initial set of
potential burn addresses. This initial set contains 208,656 addresses. We
manually classified all of them, using the following criteria.

\textbf{Burn addresses} never sent any bitcoin, have a structure that looks hand
made, and are not vanity as defined in section~\ref{subsec:background}. We used
three criteria to classify an address as a burn address: it contains words in
any language, has a repeating pattern, or is composed of an unusually small
subset of the alphabet (e.g., only numbers). We found 2,082 such addresses,
1,851 encoded in Base58 and 231 encoded in Bech32.

\textbf{Regular addresses} sent some bitcoins and do not present any hand made
structure. There are 196,088 such addresses.

\textbf{Unclassified addresses} are excluded from the training because they are
neither burn nor regular addresses. It gathers together the following categories
of addresses. i) Addresses that never sent any bitcoin and for which we observe
a structure that looks hand made, but the structure is strictly smaller than 11
characters for Base58 addresses and smaller than 13 characters for Bech32
addresses. They might be candidate burn addresses, but we cannot exclude they
are vanity addresses (see Section~\ref{subsubsec:vanity_addresses}), we found
282 such addresses. ii) Addresses that never sent any bitcoin and for which we
do not observe any structure that looks hand made. There might also be
candidates burn addresses, but here we do not observe any structure in the
address, we found 9,176 such addresses. iii) Other addresses that we do not
belong to any other categories, there are 1,028 such addresses.

In summary, we built a ground truth of 196,088 regular addresses because they all sent
some bitcoins, so they cannot be misclassified burn addresses, and of 2,082  burn
addresses because they all contain a forged structure that looks hand made and
never sent any bitcoin. In the rest of this study, we exclude all the
unclassified addresses.

\subsection{Feature selection and encoding}
\label{subsec:address_encoding}
To train a model that captures human-readable languages or
human-identifiable structures, we need an address encoding that preserves the
information about both the characters and their positions within the address.

Bitcoin addresses support two different alphabet schemes: Base58 and
Bech32~\cite{book-song,BIP173}. However, Bech32 is not a perfect subset of Base58, there
are two characters that are not present in Base58: \texttt{0} and \texttt{l} (lowercase L). We
consider the union of the Base58 and Bech32 alphabets, which we
call Base60.

To encode the frequency of characters, we use a vector $V$ of size 60 (corresponding to the length of the Base60
alphabet), where each element represents the
frequency of a character in the address. For instance, if the address contains
the letter \texttt{a} 5 times, the element at index 34 in the vector will be 5.

To encode the position, we propose a one-hot encoding represented by a matrix $M$ with
dimensions $62 \times 60$, where the rows represent the character positions in
the address (with 62 being the maximum possible address length) and the columns
represent the Base60 alphabet. Therefore, $e_{(i, j)}=1$ indicates that the
character at position $i$ of the address corresponds to the character at
position $j$ in the alphabet.

To train the model, we concatenate the vector $V$ and the matrix $M$ (flattened) to
create a feature vector of size $60 + 62 \times 60 = 3\,780$\footnote{Intuitively, it
    seems that matrix $M$ contains all information in vector $V$. However, on our tests,
    concatenating $V$ to $M$ slightly increases the number of true positive by a few
    percents, but dramatically decreases the number of false positives by one order of
    magnitude. Reducing the number of false positives is key as we need to manually check
    each address classified as burn.}.

\subsection{Model training}
\label{subsec:model_training}

We trained a Multi-Layer Perceptron (MLP) model~\cite{scikit-learn}  using the default
scikit-learn
parameters~\cite{mlp-classifier}
in version 1.5.1 (one hidden layers, 100 neurons, and the relu activation function).
To train the model, we use the dataset described in
Section~\ref{subsec:initial_training_set_construction} containing 2,082 burn addresses and the 196,088
regular addresses.

We validate the model using a stratified 10-fold cross-validation.
Table~\ref{table:cross-validation-stats} shows the excellent performance of the
model that is well adapted to our purpose. We explored the hyperparameter
space (number of hidden layers, number of neurons, solver, and activation function) and
found that the default parameters work well for our purpose. We do not
detail this exploration as it is not central to our study.

\begin{table}[t]
    \caption{Prediction and recall of the model trained to detect burn addresses using
        10-fold stratified cross validation.
        We show the mean on all folds and the standard deviation in parentheses.}
    \label{table:cross-validation-stats}
    \centering
    \begin{tabular}{|l|l|l|}
        \hline
                  & burn              & regular           \\
        \hline
        precision & 0.98825 (0.00485) & 0.99964 (0.00020) \\
        recall    & 0.96591 (0.01839) & 0.99988 (0.00005) \\
        \hline
    \end{tabular}
\end{table}

\subsection{Training, prediction, and reinforcement}
\label{subsec:reinforcement_learning}
In this section, we describe the training, prediction, and reinforcement process
we used to identify burn addresses.

First, we start by training the model on the initial training set described in
Section~\ref{subsec:initial_training_set_construction}. Then, we apply the model
on the {\maxnbaddressesnosent} addresses that never sent any bitcoin.
Indeed, as burn addresses cannot send any bitcoin, we can reduce the number of
addresses to consider from {\maxnbaddresses}, the total number of Bitcoin
addresses in our dataset, to {\maxnbaddressesnosent}.

Second, we manually scrutinize the addresses classified as burn addresses by the
model and only keep the ones that we identify as burn addresses according to the
criteria presented in Section~\ref{subsec:initial_training_set_construction}. We
add these new burn addresses to the ground truth training set and retrain the
model. In this phase, we only add burn addresses, not regular addresses. Indeed,
as the prediction is performed exclusively on addresses that never sent any
bitcoin, all addresses that we do not identify as burn addresses are considered
unclassified according to the classification in
Section~\ref{subsec:initial_training_set_construction}. Also, as the goal of the
model is to pre-filter burn addresses for manual scrutiny, we do not examine
addresses classified as regular by the model, because it is impractical due to
their large number.

Third, we iterate this training, prediction, manual scrutinization, and
reinforcement process until convergence. We present the results of this process
in Section~\ref{sec:results_model_predictions}.

\subsection{Model limitations and area of improvements}
\label{subsec:model_limitations}
The goal of this paper is to analyze the amount of bitcoins lost in burn addresses,
and the usages of burn addresses. We did not intend to build a perfect model to
detect burn addresses, but a model that is good enough to pre-filter burn
addresses for manual scrutiny. We envision several areas of improvements
for the model. First, Base58 and Bech32 addresses have a quite
different structure, mainly due to the difference in the alphabets. We could add
the address encoding as a feature to the model. Second, we used the entire address
for the training and prediction. However, the header and in particular the
trailer, which contains a checksum and always looks random, can be excluded from
the training and prediction. Last, thorough hyperparameter tuning could
improve the model performance. We did not observe any significant improvement during
our hyperparameter tuning, but we explored a limited hyperparameter space.

\subsection{Publicly available material}
\label{subsec:publicly_available_material}
We make publicly available all the material used to make
this paper. In particular, we release the initial
classification described in
Section~\ref{subsec:initial_training_set_construction} with all manual annotations, the final model trained
in Section~\ref{subsec:model_training}, the final list of burn addresses, and
the Python code to train the model and to apply the model on all Bitcoin
addresses. We provide a complete description of the material in Appendix~\ref{sec:appendix_dataset}.

\section{Results}
\label{sec:results}
\subsection{Model predictions}
\label{sec:results_model_predictions}
\begin{table}[t]
    \caption{Number of burn addresses identified at each step of the training.
        Between parenthesis, we show the new addresses identified in the step and
        never identified in all previous steps. The column \textit{initial dataset}
        shows the characteristics of the initial dataset, it does not correspond
        to a prediction, but to the initial ground truth. The other columns
        represent the prediction of the model trained on the regular addresses of
        the initial dataset and the burn addresses identified in all previous
        steps (see Section~\ref{subsec:reinforcement_learning}).}
    \label{table:training_stats}
    \centering
    \begin{tabular}{|l|l|l|l|l|}
        \hline
                            & Initial dataset & 1st prediction & 2nd prediction & 3rd prediction \\
        \hline
        \# predicted        & N/A             & 5675           & 8838           & 9672           \\
        \# true burn (TP)   & 2082            & 5376 (3294)    & 7423 (2047)    & 7905 (409)     \\
        \# false burn (FP)  & N/A             & 299            & 1415           & 1767           \\
        \hline
        \# Base58 true burn & 1851            & 3255 (1404)    & 3330 (75)      & 3346 (15)      \\
        \# Bech32 true burn & 231             & 2121 (1890)    & 4093 (1972)    & 4559 (394)     \\
        \hline
    \end{tabular}
\end{table}

We describe in the following the results of the process presented in
Section~\ref{subsec:reinforcement_learning}. Table~\ref{table:training_stats}
shows for each step of the training the number of burn addresses predicted by
the model, the number true burn addresses or true positive (TP), the number
of false burn addresses or false positive (FP), and the split of the true burn
addresses by encoding.

We observe that the number of new burn addresses identified at each step
decreases, showing a convergence of the model. However, the number of false
positives increases at each step, showing that the model tends to overfit as we
reinforce it. With a manual inspection of the false positives, we notice that
the model erroneously classifies as burn many vanity addresses. This is not
surprising as vanity addresses have a structure that looks hand made on a part
of the address, and the model is probably too simple to distinguish between burn
and vanity addresses. One way to reduce the number of false positives would be
to engineer new features that capture the difference between burn and vanity
addresses. However, this is beyond the scope of this study.

Overall, the model in the last step still has a manageable overfitting and is
good enough to pre-filter burn addresses for manual scrutiny. Out of
    {\maxnbaddressesnosent} addresses, the model extracts a set of 9,672 candidate
burn addresses. Considering the goal of the model is to pre-filter addresses on
the entire dataset for manual scrutiny, 1,767 false positives on a dataset of
    {\maxnbaddressesnosent} addresses represents 0.004\% of false positive and 18\%
of false positives on the classified addresses, an excellent result for our
purpose.

By splitting the true burn addresses by encoding, we observe that Base58
addresses are converging faster than Bech32 addresses. As the initial dataset
contains only 231 Bech32 addresses, which is 11\% of the entire initial dataset,
the model does not have enough information to identify a large fraction of the
Bech32 addresses. After the first prediction, we identified 1,890 new Bech32
burn addresses, so the model at the second prediction had enough Bech32 burn
addresses to identify a larger fraction of the Bech32 addresses. At the third
prediction, the model converges also for Bech32 addresses. In
Section~\ref{sec:results_burn_address_usages}, we will discuss in details the
difference in usages between Base58 and Bech32 burn addresses.

In summary, we created a model to automatically pre-classify burn addresses for
manual scrutiny. After a few reinforcement steps, this model identifies 7,905
true burn addresses, at the expense of a manageable number of 1,767 false
positives that can be filtered out during manual scrutinization. Finally, the model
identified 5,823 true burn addresses with a Shannon entropy greater than 4,
showing its superiority over an entropy-based classification method.

\subsection{Amount of Bitcoin lost in burn addresses}
\label{sec:results_amount_of_bitcoin_lost_in_burn_addresses}
The total amount of bitcoins lost in the 7,905 identified burn addresses is
3,197.61 bitcoins. Considering the closing price of a bitcoin on the 19th of
November 2024, 92,343~USD, this represents 295 million USD. The total supply
of bitcoins at the same date is 19.79 million bitcoins representing a fully
diluted market cap of 1.827 trillion USD. Therefore, the amount of bitcoins lost
in burn addresses represents 0.016\% of the total supply. This is a small amount
of the total supply, yet, a significant amount of USD.

\begin{table}[t]
    \caption{Quantiles of the amount of satoshi lost in each identified burn address.}
    \label{table:burn_quantiles}
    \centering
    \begin{tabular}{|l|r|r|r|r|r|}
        \hline
        quantile       & 0.50 & 0.75 & 0.90 & 0.95  & 0.99   \\
        \hline
        burned satoshi & 330  & 660  & 3000 & 10000 & 313389 \\
        \hline
    \end{tabular}
\end{table}

However, the amount of bitcoins lost in burn addresses is not evenly
distributed.
Table~\ref{table:burn_quantiles} shows the quantiles of the amount of
satoshi\footnote{A satoshi is a subdivision unit of the bitcoin. There are
    100,000,000 satoshis for one bitcoin. We use the more convenient unit depending
    on the amount considered.} burned in each identified burn address. We observe
that 75\% of the burn addresses received at most 660 satoshis, and 90\% of the
burn addresses received at most 3,000 satoshis. The Bitcoin address that burned
the largest amount of bitcoins, \texttt{1CounterpartyXXXXXXXXXXXXXXXUWLpVr},
accounts for 66.6\% of the total bitcoins lost in burn addresses. The second
largest, \texttt{1111111111111111111114oLvT2}, burned 17.6\%, and the third one,
\texttt{1ChancecoinXXXXXXXXXXXXXXXXXZELUFD}, burned 15.0\%. Together, these top
three burn addresses were responsible for over 99\% of all bitcoins lost in burn
addresses. We will discuss the specific usage of these burn addresses in
Section~\ref{sec:results_burn_address_usages}.

In summary, 0.016\% (3,197.61 bitcoins) of the total supply of bitcoins is lost in burn addresses, yet
it represents 295 million USD. The amount of bitcoins lost in burn addresses is
concentrated in a few addresses, representing different usages
of burn addresses, which we discuss in
Section~\ref{sec:results_burn_address_usages}.

\subsection{Burn address characteristics and usages}
\label{sec:results_burn_address_usages}

We show in Section~\ref{sec:results_amount_of_bitcoin_lost_in_burn_addresses}
a large inequality in the amount of bitcoins lost among burn
addresses. This inequality also represents very different usages. In the
following, we highlight representative examples of striking usages we
identified.

\subsubsection{Storing plain text messages for fun and posterity}
\label{subsubsec:plain_text_messages}
One use of burn addresses is to encode a message within a single transaction
using multiple burn addresses. In this approach, words are written directly as
plain text within the burn addresses. One striking example is illustrated by
transaction
\texttt{\href{https://mempool.space/tx/69dc9d33c39b654dc20585fe7ed848727ad8d8d04a9dd4332933cb9fad149d95}{69dc9d}}.\footnote{All
    transactions numbers are shortened and hyperlink. The complete Tx numbers are in Appendix~\ref{sec:appendix_referenced_transactions}.}.
It has been forged in July 2021 and encoded with 424 burn addresses a message to
advocate for the usage of proof-of-stake in Bitcoin. This is the largest message
encoded in Base58 burn addresses we found. This message also contains two
addresses to vote for or against the proposal, however, nobody every voted to
these addresses. We found 178 addresses in the initial set, our model
automatically found 243 more, only 3 were missed. We provide additional examples
of this use of burn addresses in
Annex~\ref{subsec:appendix_message_encoded_in_multiple_burn_addresses}.

Another frequent use of burn addresses is to encode a short message within a
single address. Indeed, our analysis revealed that 60\% of the identified Base58
burn addresses appear in transactions without any other burn addresses. Examples
of this use are \texttt{1iSeeYouMrRobotXXVeryNiceXXVwyoA6} in transaction
\texttt{\href{https://mempool.space/tx/f784ede1963d2f83d087f76b62a94896aba0aef9df1e89a6c7064f47220f4b43}{f784ed}}
or \texttt{1HeyYouGetBackToWorkNowN2b8UHSeTzi} in transaction
\texttt{\href{https://mempool.space/tx/4659082fbd227a84a53c748c2519364148418213c33d7b584da4998be1b53cb3}{465908}}.
We give more examples in
Annex~\ref{subsec:appendix_message_encoded_in_single_burn_address}.

\subsubsection{Storing images for fun and posterity}
\label{subsubsec:images}
We did not observe plain text messages encoded in Bech32 addresses. As our
methodology is designed to identify burn addresses with a human-readable
structure or easily identifiable patterns, we are not supposed to detect images
encoded in burn addresses. However, images might contain uniform patterns (such
as a monochromatic background) or padding, which might result in identifiable
patterns in addresses. For instance, padding is often performed with zeros, and
in Bech32 encoding, 0 is represented as the letter \texttt{q}. Therefore, in the initial dataset, we
identified 231 Bech32 addresses containing a lot of \texttt{q} characters as candidate
burn addresses. Then, the model identified 4,559 Bech32 burn addresses
containing repeating patterns, still with no human-readable text.

We observed that Bech32 burn addresses had a sudden increase in usage early 2024
(the ECDF of burn addresses apparition date is available in
Appendix~\ref{subsec:appendix_message_encoded_in_multiple_burn_addresses} in
Fig.~\ref{fig:burn_oldness_ecdf}). These addresses are not part of a
proof-of-burn protocol, but part of new protocols introduced early 2024: OLGA
Stamps~\cite{stamps-olga} leverages on the Counterparty
protocol~\cite{counterparty} to store images in the blockchain, the
SRC-20~\cite{src20specs} and SRC-721~\cite{src721-repository} token
specifications store tokens directly on Bitcoin using the Bitcoin Stamps
protocol~\cite{stamps-sdk}. Using a Counterparty Decoder developed by J.P.
Janssen~\cite{electrum-counterparty-decode-tx}, we were able to extract the
images encoded in the Bech32 burn addresses.

For instance, address
\texttt{\href{https://mempool.space/address/bc1qx56r2dgqy8usgpg9qqqsqtqpqq9qq8sqqyqqqqs9sj86j6c9qqssptq452}{bc1qx56r2}}
(see Annex~\ref{sec:appendix_referenced_addresses}) is in transaction
\texttt{\href{https://mempool.space/tx/926920c4adce5f3fa2171ee4be337c79eb1aa580295bf7ea38e1a52e2276f613}{926920}}
forged in March 2024 using OLGA Stamps and 100 burn addresses. It represents an
animated GIF of an abstract subject. None of the burn address were identified in
the initial set, the model identified 16 burn addresses, 84 were missed. We miss
many of these burn addresses, because they do not contain any human identifiable
pattern.

Interestingly, we also found images encoded in Base58 burn addresses. For instance, we identified address
\texttt{111111111111111K3tBycEZAhc5M} that is part of transaction
\texttt{\href{https://mempool.space/tx/6240f61bbaeac66cd623e921a153addaf5f379a996f2de0f0c6506d628fe3812}{6240f6}},
which contains the Superbuffo image~\cite{Santilli2024}

\subsubsection{Proof-of-burn}
\label{subsubsec:proof_of_burn}
From our results, proof-of-burn is by far the largest source of burned bitcoins,
but it has been successfully applied by only few projects. In 2014, the Counterparty
project~\cite{counterparty} used the concept of
proof-of-burn~\cite{stewart-proof-of-burn} for the creation of their XCP token
created by burning bitcoins at the address
\texttt{1CounterpartyXXXXXXXXXXXXXXXUWLpVr}~\cite{proof-of-burn-article}. The
same year, the Chancecoin protocol also used proof-of-burn to create their CHA
token by burning bitcoins at the address
\texttt{1ChancecoinXXXXXXXXXXXXXXXXXZELUFD}~\cite{chancecoin}.

\subsubsection{Protocol operation}
Blockstack~\cite{blockstack2016} proposes a global naming and storage system on
top of other blockchains, and uses the burn address
\texttt{1111111111111111111114oLvT2} in its name space registration
process~\cite{blockstack-wire-format}. This burn address is not used to create a
new token, but to permanently store a state required by Blockstack in the
Bitcoin blockchain. This address corresponds to an address containing only zeros when
expressed in hexadecimal. Therefore, it cannot be excluded that this address has been
used for other purposes.

\section{Related work}
\label{sec:related_work}
In 2012, Iain Stewart wrote an early description of the notion of
proof-of-burn~\cite{stewart-proof-of-burn} describing possible usages including the creation
of a new cryptocurrency.
This last usage has been demonstrated, in 2014, by the Counterparty project for
the creation of its XCP token created by burning bitcoins at the address
\texttt{1CounterpartyXXXXXXXXXXXXXXXUWLpVr}~\cite{proof-of-burn-article}.
Karantias et al.~\cite{Karantias2020} defined the notion of proof-of-burn as a
cryptographic primitive with unspendability, binding, and uncensorability
properties.

Some literature explores using Bitcoin to persist arbitrary data unrelated to
bitcoin creation or exchange
\cite{Bartoletti2019,matzutt2018quantitative,scheid2023opening,sward2018data}.
Bartoletti et al.~\cite{Bartoletti2019}  identified 11 different possibilities
to embed metadata in the blockchain, the most popular one being the usage of
\texttt{OP\_RETURN} transactions. Matzutt et al.~\cite{matzutt2018quantitative}
observed objectionable content in the Bitcoin blockchain and raise interesting
questions about the liability people downloading the blockchain could face. More
recently, Scheid et al.~\cite{scheid2023opening} studied the storage of
arbitrary data in Bitcoin, Monero, and Ethereum.
All these works cite the notion of burn addresses as a consequence of storing
data in transactions that cannot be spent. However, they do not specifically
study burn addresses and do not quantify their usage.

Sward et al.~\cite{sward2018data} explore how arbitrary data is stored in the
Bitcoin blockchain, in particular, they describe methods to store arbitrary data
with the consequence of burning bitcoins. However, their technique is
much simpler and less reliable than ours as it consists in finding a given
threshold of ASCII characters in the addresses for P2PKH UTXOs. Bistarelli et
al.~\cite{bistarelli2019analysis} specifically focused on non-standard
transactions. However, they do not specifically study burn addresses
(non-standard transactions do not necessarily use burn addresses) and do not
quantify their usage.
Ciro Santilli~\cite{Santilli2024} maintains a Web page listing arbitrary data
embedded in the Bitcoin blockchain. This is the most complete and up-to-date
list of arbitrary data we are aware of. However, he does not specifically focus
on burn addresses, but on the stored contents.

To the best of our knowledge, this paper is the first one to propose a
methodology for identifying burn addresses, quantifying the amount of bitcoins
lost in burn addresses, analyzing their uses, and releasing a dataset of 7,905
burn addresses and a trained model to identify new burn addresses.

\section{Conclusion}
\label{sec:conclusion}
In this paper, we proposed a methodology to identify burn addresses, quantified
the permanent losses of bitcoins, and explored the uses of burn address.

Although the practice of burning bitcoins is criticized, it is \textit{de facto}
used. Bitcoin is an ideal candidate for burning coins for two main
reasons: first, bitcoins have the highest value among cryptocurrencies, making
them perfect candidate to burn to create new tokens; and second, the Bitcoin
blockchain is highly secure and immutable, making it a perfect candidate to
store data permanently. Indeed, we observed that Bitcoin burn addresses are used to
create new tokens with proof-of-burn, to store plain text messages or images,
and to operate protocols.
We believe that this paper sheds a new light on the uses of Bitcoin burn
addresses and paves the way for further research on the topic by making
public the ground truth training set, the list of identified burn addresses, and
the model to identify new burn addresses.

\newpage
\bibliographystyle{ACM-Reference-Format}
\bibliography{bibliography}


\begin{thebibliography}{30}


\ifx \showCODEN    \undefined \def \showCODEN     #1{\unskip}     \fi
\ifx \showDOI      \undefined \def \showDOI       #1{#1}\fi
\ifx \showISBNx    \undefined \def \showISBNx     #1{\unskip}     \fi
\ifx \showISBNxiii \undefined \def \showISBNxiii  #1{\unskip}     \fi
\ifx \showISSN     \undefined \def \showISSN      #1{\unskip}     \fi
\ifx \showLCCN     \undefined \def \showLCCN      #1{\unskip}     \fi
\ifx \shownote     \undefined \def \shownote      #1{#1}          \fi
\ifx \showarticletitle \undefined \def \showarticletitle #1{#1}   \fi
\ifx \showURL      \undefined \def \showURL       {\relax}        \fi
\providecommand\bibfield[2]{#2}
\providecommand\bibinfo[2]{#2}
\providecommand\natexlab[1]{#1}
\providecommand\showeprint[2][]{arXiv:#2}

\bibitem[pro(2014)]%
        {proof-of-burn-article}
 \bibinfo{year}{2014}\natexlab{}.
\newblock \bibinfo{booktitle}{\emph{Why Proof of Burn?}}
\newblock
\urldef\tempurl%
\url{https://www.counterparty.io/news/why-proof-of-burn/}
\showURL{%
\tempurl}
\newblock
\shownote{Last accessed: 2024-09-24}.


\bibitem[cha(2024)]%
        {chancecoin}
 \bibinfo{year}{2024}\natexlab{}.
\newblock \bibinfo{booktitle}{\emph{Chancecoin}}.
\newblock
\urldef\tempurl%
\url{https://bitcointalk.org/index.php?topic=528023.0}
\showURL{%
\tempurl}
\newblock
\shownote{Accessed: 2024-11-21}.


\bibitem[cou(2024)]%
        {counterparty}
 \bibinfo{year}{2024}\natexlab{}.
\newblock \bibinfo{booktitle}{\emph{Counterparty}}.
\newblock
\urldef\tempurl%
\url{https://www.counterparty.io/}
\showURL{%
\tempurl}
\newblock
\shownote{Accessed: 2024-11-21}.


\bibitem[blo(2024)]%
        {blockstack-wire-format}
 \bibinfo{year}{2024}\natexlab{}.
\newblock \bibinfo{booktitle}{\emph{Creation of name operation in blockstack}}.
\newblock
\urldef\tempurl%
\url{https://github.com/stacks-network/stacks-core/blob/647234c4b2aa11a976d6e405e3e4d9fa69497456/docs/wire-format.md}
\showURL{%
\tempurl}
\newblock
\shownote{Accessed: 2024-11-21}.


\bibitem[mlp(2024)]%
        {mlp-classifier}
 \bibinfo{year}{2024}\natexlab{}.
\newblock \bibinfo{title}{MLP Classifier in scikit-learn}.
\newblock
  \bibinfo{howpublished}{\url{https://scikit-learn.org/stable/modules/generated/sklearn.neural_network.MLPClassifier.html}}.
\newblock


\bibitem[Ali et~al\mbox{.}(2016)]%
        {blockstack2016}
\bibfield{author}{\bibinfo{person}{Muneeb Ali}, \bibinfo{person}{Jude Nelson},
  \bibinfo{person}{Ryan Shea}, {and} \bibinfo{person}{Michael~J. Freedman}.}
  \bibinfo{year}{2016}\natexlab{}.
\newblock \showarticletitle{Blockstack: A Global Naming and Storage System
  Secured by Blockchains}. In \bibinfo{booktitle}{\emph{2016 USENIX Annual
  Technical Conference (USENIX ATC 16)}}. \bibinfo{publisher}{USENIX
  Association}, \bibinfo{address}{Denver, CO}, \bibinfo{pages}{181--194}.
\newblock


\bibitem[Andreas M~Antonopoulos(2023)]%
        {book-mastering-btc}
\bibfield{author}{\bibinfo{person}{David A~Harding Andreas M~Antonopoulos}.}
  \bibinfo{year}{2023}\natexlab{}.
\newblock \bibinfo{booktitle}{\emph{Mastering Bitcoin: Programming the Open
  Blockchain}}.
\newblock \bibinfo{publisher}{O'Reilly Media}.
\newblock
\showISBNx{978-1098150099}


\bibitem[Bartoletti et~al\mbox{.}(2019)]%
        {Bartoletti2019}
\bibfield{author}{\bibinfo{person}{Massimo Bartoletti}, \bibinfo{person}{Bryn
  Bellomy}, {and} \bibinfo{person}{Livio Pompianu}.}
  \bibinfo{year}{2019}\natexlab{}.
\newblock \showarticletitle{A journey into bitcoin metadata}.
\newblock \bibinfo{journal}{\emph{Journal of Grid Computing}}
  \bibinfo{volume}{17} (\bibinfo{year}{2019}), \bibinfo{pages}{3--22}.
\newblock


\bibitem[Bistarelli et~al\mbox{.}(2019)]%
        {bistarelli2019analysis}
\bibfield{author}{\bibinfo{person}{Stefano Bistarelli}, \bibinfo{person}{Ivan
  Mercanti}, {and} \bibinfo{person}{Francesco Santini}.}
  \bibinfo{year}{2019}\natexlab{}.
\newblock \showarticletitle{An analysis of non-standard transactions}.
\newblock \bibinfo{journal}{\emph{Frontiers in Blockchain}}
  \bibinfo{volume}{2} (\bibinfo{year}{2019}), \bibinfo{pages}{7}.
\newblock


\bibitem[contributors(2023)]%
        {emeraldtablet-wikipedia}
\bibfield{author}{\bibinfo{person}{Wikipedia contributors}.}
  \bibinfo{year}{2023}\natexlab{}.
\newblock \bibinfo{title}{Emerald Tablet}.
\newblock
\newblock
\urldef\tempurl%
\url{https://en.wikipedia.org/wiki/Emerald_Tablet}
\showURL{%
\tempurl}
\newblock
\shownote{Accessed: 2024-10-01}.


\bibitem[contributors(2024)]%
        {camelCase-wikipedia}
\bibfield{author}{\bibinfo{person}{Wikipedia contributors}.}
  \bibinfo{year}{2024}\natexlab{}.
\newblock \bibinfo{title}{CamelCase}.
\newblock
\newblock
\urldef\tempurl%
\url{https://en.wikipedia.org/wiki/Camel_case}
\showURL{%
\tempurl}
\newblock
\shownote{Accessed: 2024-11-20}.


\bibitem[DerpHerpenstein(nd)]%
        {src721-repository}
\bibfield{author}{\bibinfo{person}{DerpHerpenstein}.}
  \bibinfo{year}{n.d.}\natexlab{}.
\newblock \bibinfo{title}{SRC-721 Specification}.
\newblock
  \bibinfo{howpublished}{\url{https://github.com/DerpHerpenstein/src-721}}.
\newblock
\newblock
\shownote{Accessed: 2024-10-01}.


\bibitem[Janssen(nd)]%
        {electrum-counterparty-decode-tx}
\bibfield{author}{\bibinfo{person}{J.P. Janssen}.}
  \bibinfo{year}{n.d.}\natexlab{}.
\newblock \bibinfo{title}{Electrum-Counterparty: Decode Transaction}.
\newblock
  \bibinfo{howpublished}{\url{https://jpja.github.io/Electrum-Counterparty/decode_tx}}.
\newblock
\newblock
\shownote{Accessed: 2024-10-01}.


\bibitem[Karantias et~al\mbox{.}(2020)]%
        {Karantias2020}
\bibfield{author}{\bibinfo{person}{Kostis Karantias}, \bibinfo{person}{Aggelos
  Kiayias}, {and} \bibinfo{person}{Dionysis Zindros}.}
  \bibinfo{year}{2020}\natexlab{}.
\newblock \showarticletitle{Proof-of-Burn}. In
  \bibinfo{booktitle}{\emph{Financial Cryptography and Data Security}}.
  \bibinfo{pages}{523--540}.
\newblock


\bibitem[Kiss(2023)]%
        {pycoin}
\bibfield{author}{\bibinfo{person}{Richard Kiss}.}
  \bibinfo{year}{2023}\natexlab{}.
\newblock \bibinfo{title}{pycoin -- Python Cryptocoin Utilities}.
\newblock
\newblock
\urldef\tempurl%
\url{https://github.com/richardkiss/pycoin}
\showURL{%
\tempurl}
\newblock
\shownote{Version 0.92.20230326}.


\bibitem[Matzutt et~al\mbox{.}(2018)]%
        {matzutt2018quantitative}
\bibfield{author}{\bibinfo{person}{Roman Matzutt}, \bibinfo{person}{Jens
  Hiller}, \bibinfo{person}{Martin Henze}, \bibinfo{person}{Jan~Henrik
  Ziegeldorf}, \bibinfo{person}{Dirk M{\"u}llmann}, \bibinfo{person}{Oliver
  Hohlfeld}, {and} \bibinfo{person}{Klaus Wehrle}.}
  \bibinfo{year}{2018}\natexlab{}.
\newblock \showarticletitle{A quantitative analysis of the impact of arbitrary
  blockchain content on bitcoin}. In \bibinfo{booktitle}{\emph{Financial
  Cryptography and Data Security}}. Springer, \bibinfo{pages}{420--438}.
\newblock


\bibitem[Mikeinspace(2024)]%
        {stamps-olga}
\bibfield{author}{\bibinfo{person}{Mikeinspace}.}
  \bibinfo{year}{2024}\natexlab{}.
\newblock \bibinfo{title}{Bitcoin Stamps Improvement Proposal: Octet Linked
  Graphics \& Artifacts (OLGA)}.
\newblock
  \bibinfo{howpublished}{\url{https://github.com/stampchain-io/stamps/blob/main/OLGA.md}}.
\newblock
\newblock
\shownote{Accessed: 2024-10-01}.


\bibitem[Nakamoto(2009)]%
        {nakamoto2009bitcoin}
\bibfield{author}{\bibinfo{person}{Satoshi Nakamoto}.}
  \bibinfo{year}{2009}\natexlab{}.
\newblock \bibinfo{title}{Bitcoin: A Peer-to-Peer Electronic Cash System}.
  (\bibinfo{date}{May} \bibinfo{year}{2009}).
\newblock
\urldef\tempurl%
\url{http://www.bitcoin.org/bitcoin.pdf}
\showURL{%
\tempurl}


\bibitem[Nakamoto and the Bitcoin Core~contributors(2009)]%
        {bitcoin-core}
\bibfield{author}{\bibinfo{person}{Satoshi Nakamoto} {and} \bibinfo{person}{the
  Bitcoin Core~contributors}.} \bibinfo{year}{2009}\natexlab{}.
\newblock \bibinfo{title}{Bitcoin Core: A peer-to-peer electronic cash system}.
\newblock
\newblock
\urldef\tempurl%
\url{https://bitcoincore.org/}
\showURL{%
\tempurl}
\newblock
\shownote{Version 27.1}.


\bibitem[Pedregosa et~al\mbox{.}(2011)]%
        {scikit-learn}
\bibfield{author}{\bibinfo{person}{F. Pedregosa}, \bibinfo{person}{G.
  Varoquaux}, \bibinfo{person}{A. Gramfort}, \bibinfo{person}{V. Michel},
  \bibinfo{person}{B. Thirion}, \bibinfo{person}{O. Grisel},
  \bibinfo{person}{M. Blondel}, \bibinfo{person}{P. Prettenhofer},
  \bibinfo{person}{R. Weiss}, \bibinfo{person}{V. Dubourg}, \bibinfo{person}{J.
  Vanderplas}, \bibinfo{person}{A. Passos}, \bibinfo{person}{D. Cournapeau},
  \bibinfo{person}{M. Brucher}, \bibinfo{person}{M. Perrot}, {and}
  \bibinfo{person}{E. Duchesnay}.} \bibinfo{year}{2011}\natexlab{}.
\newblock \showarticletitle{Scikit-learn: Machine Learning in {P}ython}.
\newblock \bibinfo{journal}{\emph{Journal of Machine Learning Research}}
  \bibinfo{volume}{12} (\bibinfo{year}{2011}), \bibinfo{pages}{2825--2830}.
\newblock


\bibitem[Santilli(2024)]%
        {Santilli2024}
\bibfield{author}{\bibinfo{person}{Ciro Santilli}.}
  \bibinfo{year}{2024}\natexlab{}.
\newblock \bibinfo{title}{Cool data embedded in the Bitcoin blockchain}.
\newblock
\newblock
\urldef\tempurl%
\url{https://cirosantilli.com/cool-data-embedded-in-the-bitcoin-blockchain}
\showURL{%
\tempurl}
\newblock
\shownote{Accessed: 2024-09-22}.


\bibitem[Scheid et~al\mbox{.}(2023)]%
        {scheid2023opening}
\bibfield{author}{\bibinfo{person}{Eder~J Scheid}, \bibinfo{person}{Sebastian
  K{\"u}ng}, \bibinfo{person}{Muriel~F Franco}, {and} \bibinfo{person}{Burkhard
  Stiller}.} \bibinfo{year}{2023}\natexlab{}.
\newblock \showarticletitle{Opening Pandora's Box: An Analysis of the Usage of
  the Data Field in Blockchains}. In \bibinfo{booktitle}{\emph{Fifth
  International Conference on Blockchain Computing and Applications}}. IEEE,
  \bibinfo{pages}{369--376}.
\newblock


\bibitem[Shannon(1948)]%
        {shannon-entropy}
\bibfield{author}{\bibinfo{person}{C.~E. Shannon}.}
  \bibinfo{year}{1948}\natexlab{}.
\newblock \showarticletitle{A mathematical theory of communication}.
\newblock \bibinfo{journal}{\emph{The Bell System Technical Journal}}
  \bibinfo{volume}{27}, \bibinfo{number}{3} (\bibinfo{year}{1948}),
  \bibinfo{pages}{379--423}.
\newblock
\urldef\tempurl%
\url{https://doi.org/10.1002/j.1538-7305.1948.tb01338.x}
\showDOI{\tempurl}


\bibitem[Song(2019)]%
        {book-song}
\bibfield{author}{\bibinfo{person}{Jimmy Song}.}
  \bibinfo{year}{2019}\natexlab{}.
\newblock \bibinfo{booktitle}{\emph{Programming Bitcoin: Learn How to Program
  Bitcoin from Scratch} (\bibinfo{edition}{1st} ed.)}.
\newblock \bibinfo{publisher}{O'Reilly Media, Inc.}
\newblock
\showISBNx{1492031496}


\bibitem[Stampchain.io(nda)]%
        {src20specs}
\bibfield{author}{\bibinfo{person}{Stampchain.io}.}
  \bibinfo{year}{n.d.}\natexlab{a}.
\newblock \bibinfo{title}{SRC20 Specification Documentation}.
\newblock
  \bibinfo{howpublished}{\url{https://github.com/stampchain-io/stamps_sdk/blob/main/docs/src20specs.md}}.
\newblock
\newblock
\shownote{Accessed: 2024-10-01}.


\bibitem[Stampchain.io(ndb)]%
        {stamps-sdk}
\bibfield{author}{\bibinfo{person}{Stampchain.io}.}
  \bibinfo{year}{n.d.}\natexlab{b}.
\newblock \bibinfo{title}{Stamps SDK}.
\newblock
  \bibinfo{howpublished}{\url{https://github.com/stampchain-io/stamps_sdk/tree/main}}.
\newblock
\newblock
\shownote{Accessed: 2024-10-01}.


\bibitem[Stewart(2012)]%
        {stewart-proof-of-burn}
\bibfield{author}{\bibinfo{person}{Iain Stewart}.}
  \bibinfo{year}{2012}\natexlab{}.
\newblock \bibinfo{booktitle}{\emph{Proof of Burn}}.
\newblock
\urldef\tempurl%
\url{https://en.bitcoin.it/wiki/Proof_of_burn}
\showURL{%
\tempurl}
\newblock
\shownote{Last accessed: 2024-09-24}.


\bibitem[Sward et~al\mbox{.}(2018)]%
        {sward2018data}
\bibfield{author}{\bibinfo{person}{Andrew Sward}, \bibinfo{person}{Ivy Vecna},
  {and} \bibinfo{person}{Forrest Stonedahl}.} \bibinfo{year}{2018}\natexlab{}.
\newblock \showarticletitle{Data insertion in Bitcoin's blockchain}.
\newblock \bibinfo{journal}{\emph{Ledger}}  \bibinfo{volume}{3}
  (\bibinfo{year}{2018}).
\newblock


\bibitem[{Wikipedia contributors}(2024)]%
        {aristocrats-wikipedia}
\bibfield{author}{\bibinfo{person}{{Wikipedia contributors}}.}
  \bibinfo{year}{2024}\natexlab{}.
\newblock \bibinfo{title}{The Aristocrats}.
\newblock
\newblock
\urldef\tempurl%
\url{https://en.wikipedia.org/wiki/The_Aristocrats}
\showURL{%
\tempurl}
\newblock
\shownote{Accessed: 2024-10-01}.


\bibitem[Wuille and Maxwell(2017)]%
        {BIP173}
\bibfield{author}{\bibinfo{person}{Pieter Wuille} {and} \bibinfo{person}{Greg
  Maxwell}.} \bibinfo{year}{2017}\natexlab{}.
\newblock \bibinfo{title}{Base32 address format for native v0-16 witness
  outputs}.
\newblock \bibinfo{howpublished}{Bitcoin Improvement Proposal}.
\newblock
\urldef\tempurl%
\url{https://github.com/bitcoin/bips/blob/master/bip-0173.mediawiki}
\showURL{%
\tempurl}
\newblock
\shownote{BIP 173, Final, Informational}.


\end{thebibliography}

\newpage
\appendix

\section{Acknowledgments}
The authors want to thank Vojtěch Strnad on
\href{https://bitcoin.stackexchange.com}{bitcoin.stackexchange.com} for help
in identifying the OLGA Stamp protocol and decoding the stored images, and Damien Saucez for suggestions on the manuscript.

\section{Ethics}
This work does not raise any ethical issues

\section{Declaration of generative AI and AI-assisted technologies in the writing process }
During the preparation of this work the authors used ChatGPT (models GPT-4o and
GPT-4o mini)  to improve language. The use was limited to sentence fragments.
All propositions were carefully reviewed, modified, and integrated by the
authors. We never used a generative AI to analyze results and write full
sentences or paragraphs. The authors take full responsibility for the content of
the publication.

\section{Histogram of the Shannon entropy for all
  burn addresses}
\label{sec:appendix_histogram_shannon_entropy}
\begin{figure}[h]
    \centering
    \includegraphics[width=0.6\columnwidth]{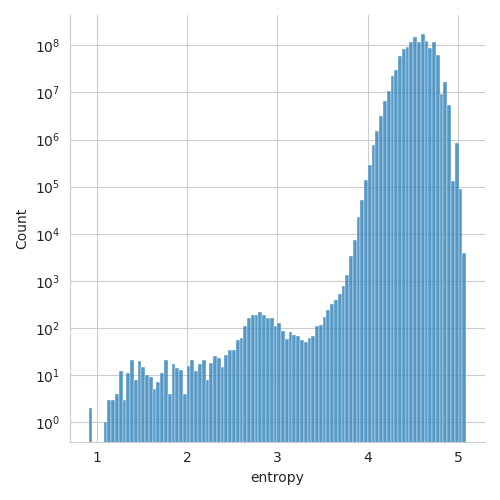}
    \caption{Histogram of the Shannon entropy for all Bitcoin addresses. The y-axis is log scale.}
    \label{fig:entropy_hist}
\end{figure}

\section{Storing plain text messages for fun and posterity}
\label{sec:appendix_plain_text_messages}
In this section, we present additional examples of messages encoded in burn
addresses. We do not intend to be exhaustive, but to illustrate the variety of
messages encoded.

\subsection{Message encoded in multiple burn addresses}
\label{subsec:appendix_message_encoded_in_multiple_burn_addresses}
\begin{figure}[h]
    \centering
    \includegraphics[width=0.6\columnwidth]{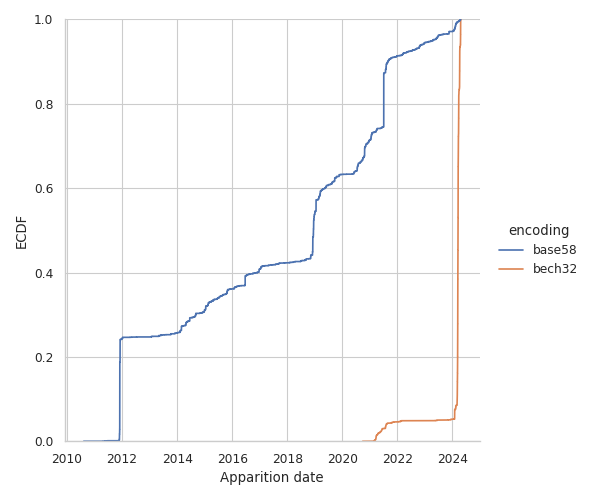}
    \caption{Distribution of the first apparition date of each burn address.}
    \label{fig:burn_oldness_ecdf}
\end{figure}

We see in Figure~\ref{fig:burn_oldness_ecdf} that there are three peaks in the
creation of Base58 burn addresses in late 2011, late 2018, and mid-2021. Each of
these events involves encoding plain text messages, with each message encoded in
a single transaction using multiple burn addresses. In the following, we give a
few striking examples of these messages.

Transaction \texttt{\href{https://mempool.space/tx/3bbd94d22346a3bfb44257293e10c3b5c9ee39230c1cd358bdce2bf03c61ba0b}{3bbd94}}
has been forged in November 2011 and encoded with  49 burn addresses a
modified version of the text on the Emerald
Tablet~\cite{emeraldtablet-wikipedia}. We identified 24 addresses in the
initial set, our model automatically found 17 more, but 8 were missed. Still
in November 2011, Transaction
\texttt{\href{https://mempool.space/tx/28ccf29cfcc9f82d42793db770e7c7894d61ccf3d18299f34bda2e54415da287}{28ccf2}}
encoded with 8 addresses a small excerpt of \textit{Alice's
    Adventures in Wonderland} by Lewis Carroll. We identified four addresses in
the initial set, our model automatically found three more, and one was
missed. Transaction
\texttt{\href{https://mempool.space/tx/8ffacbb18f63576fe323cbf2acc6c4c01c86aadf13d8352cfdd39d91916d98c8}{8ffacb}}
forged in December 2011 encoded with 240 burn addresses a repeating small
advertisement claiming to give blockchain immortality with a service called
\texttt{etchablock.com}. Interestingly, the issuer of this transaction is
the address
\href{https://mempool.space/address/1DKMg2KmTyQmfiQxDfDsNKVx4PWYa7xga4}{1DKMg2K}
that also issued more than 30 transactions between December 1st and December 7th
2011 encoding messages with burn addresses. Most likely, it was an attempt to test
a service to encode messages with burn addresses.
Transaction
\texttt{\href{https://mempool.space/tx/028b8514a4f6cc96ac3c1c83dbb117ab9dc5eb09deab7b49bf038fd460173127}{028b85}}
forged in December 2011 encoded with 81 burn addresses an excerpt of the Aristocrats
joke~\cite{aristocrats-wikipedia}. We identified 24 addresses in the initial
set, our model automatically found 570 more, and none was missed.

Transaction
\texttt{\href{https://mempool.space/tx/57bfd63000bbfa6e9a61f7285a4abf9aef91dfcfba4fe0f940b431653eb8068b}{57bfd6}}
forged in December 2018 encoded with 270 burn addresses a variation of the
classic "Lorem Ipsum" placeholder text. We identified 53 addresses in the
initial set, our model automatically found 134 more, 83 were missed.
Transaction
\texttt{\href{https://mempool.space/tx/e9a418b6d8e6b98f465ac9ce2bc92c16069bb16e48d68b90f8c131bbb3742675}{e9a418}}
forged in December 2018 encoded with 74 addresses the
lyrics of the song "Too Late, Game Over \& Goodbye" by Krypteria, a German gothic metal band.

The peaks in mid-2021 are due to the transaction
Transaction \texttt{\href{https://mempool.space/tx/69dc9d33c39b654dc20585fe7ed848727ad8d8d04a9dd4332933cb9fad149d95}{69dc9d}}
that is described in Section~\ref{subsubsec:plain_text_messages}.

\subsection{Message encoded in a single burn address}
\label{subsec:appendix_message_encoded_in_single_burn_address}
In this section, we present examples of messages encoded in a single burn
address.
\\\\
\Small
\noindent
\texttt{1KToEddieAndELLieLoveYouDadXWz9SPM} in transaction
\texttt{\href{https://mempool.space/tx/9508f0f37aa91abe92b66fc42a43ca801a7b4eca5d246fb7dd73db8eb91ad04e}{9508f0}}.\\
\texttt{17MarHappyBirthdayWuYunChengUSerRi} in transaction
\texttt{\href{https://mempool.space/fr/tx/af7e82e4ec88bbda055a4ee9319290635e7cac1a4d45f131d1182a5922a7d22b}{af7e82}}.\\
\texttt{1CrackedBitcoinHaHaHaHaHaHaHUvfWst} in transaction \texttt{\href{https://mempool.space/fr/tx/46e876485226b620e11a4d64bbbff92d9f0fd63161d8e68de0c13252f8187dc8}{46e876}}\\
\texttt{1GodB1essAmericaxxxxxxxxxxxyEGQkE} in transaction \texttt{\href{https://mempool.space/fr/tx/ae70f72a454ef7016c828db6d33a92550c797958f73856a220eae23b93da0588}{ae70f7}}\\
\texttt{1HELP1ME1PLEASExxxxxxxxxxxxzn42ey} in transaction \texttt{\href{https://mempool.space/fr/tx/9b9fd0150a1249c8c4e403b41f47070f52e1d0538bfc422b1684febe50cda103}{9b9fd0}}\\
\texttt{1HappyBirthDayToYou1Martica11cNVNe} in transaction \texttt{\href{https://mempool.space/fr/tx/b5853bb7ce02e915aa8eac59b1256f1599247b27872b0b0786550c580dead652}{b5853b}}\\
\texttt{1HeyDanThanksALotForYourTimeWeBkbn} in transaction \texttt{\href{https://mempool.space/fr/tx/4a13a422d8da493664be515fd8175775b77fe04aee599695275e5ecd6fae720f}{4a13a4}}\\
\texttt{1JeSuisChar1ieXXXXXXXXXXXXXXZCQPGy} in transaction \texttt{\href{https://mempool.space/fr/tx/7dcd0538ad7bdcdb79d7d0244cf4ff002b46779e5594fe8b84fd1939d9770999}{7dcd05}}\\
\texttt{1LoveAndMissYouSarahSwainXXXSmYjR7} in transaction \texttt{\href{https://mempool.space/fr/tx/c82f14c3e1bdf9651dc1f7521d8624586c3d4c603dd03dc43890531ba41da0ac}{c82f14}}\\
\normalsize

\section{Referenced transactions}
\label{sec:appendix_referenced_transactions}
We present in this section the complete transaction numbers of the transactions
referenced in the main text and in the appendices. All Tx are hyperlinked to
\url{https://mempool.space}.
\\\\
\Small 
\noindent
\href{https://mempool.space/tx/69dc9d33c39b654dc20585fe7ed848727ad8d8d04a9dd4332933cb9fad149d95}{69dc9d33c39b654dc20585fe7ed848727ad8d8d04a9dd4332933cb9fad149d95}\\
\href{https://mempool.space/tx/f784ede1963d2f83d087f76b62a94896aba0aef9df1e89a6c7064f47220f4b43}{f784ede1963d2f83d087f76b62a94896aba0aef9df1e89a6c7064f47220f4b43}\\
\href{https://mempool.space/tx/4659082fbd227a84a53c748c2519364148418213c33d7b584da4998be1b53cb3}{4659082fbd227a84a53c748c2519364148418213c33d7b584da4998be1b53cb3}\\
\href{https://mempool.space/tx/9508f0f37aa91abe92b66fc42a43ca801a7b4eca5d246fb7dd73db8eb91ad04e}{9508f0f37aa91abe92b66fc42a43ca801a7b4eca5d246fb7dd73db8eb91ad04e}\\
\href{https://mempool.space/tx/926920c4adce5f3fa2171ee4be337c79eb1aa580295bf7ea38e1a52e2276f613}{926920c4adce5f3fa2171ee4be337c79eb1aa580295bf7ea38e1a52e2276f613}\\
\href{https://mempool.space/tx/6240f61bbaeac66cd623e921a153addaf5f379a996f2de0f0c6506d628fe3812}{6240f61bbaeac66cd623e921a153addaf5f379a996f2de0f0c6506d628fe3812}\\
\href{https://mempool.space/tx/3bbd94d22346a3bfb44257293e10c3b5c9ee39230c1cd358bdce2bf03c61ba0b}{3bbd94d22346a3bfb44257293e10c3b5c9ee39230c1cd358bdce2bf03c61ba0b}\\
\href{https://mempool.space/tx/28ccf29cfcc9f82d42793db770e7c7894d61ccf3d18299f34bda2e54415da287}{28ccf29cfcc9f82d42793db770e7c7894d61ccf3d18299f34bda2e54415da287}\\
\href{https://mempool.space/tx/8ffacbb18f63576fe323cbf2acc6c4c01c86aadf13d8352cfdd39d91916d98c8}{8ffacbb18f63576fe323cbf2acc6c4c01c86aadf13d8352cfdd39d91916d98c8}\\
\href{https://mempool.space/tx/028b8514a4f6cc96ac3c1c83dbb117ab9dc5eb09deab7b49bf038fd460173127}{028b8514a4f6cc96ac3c1c83dbb117ab9dc5eb09deab7b49bf038fd460173127}\\
\href{https://mempool.space/tx/57bfd63000bbfa6e9a61f7285a4abf9aef91dfcfba4fe0f940b431653eb8068b}{57bfd63000bbfa6e9a61f7285a4abf9aef91dfcfba4fe0f940b431653eb8068b}\\
\href{https://mempool.space/tx/e9a418b6d8e6b98f465ac9ce2bc92c16069bb16e48d68b90f8c131bbb3742675}{e9a418b6d8e6b98f465ac9ce2bc92c16069bb16e48d68b90f8c131bbb3742675}\\
\href{https://mempool.space/tx/69dc9d33c39b654dc20585fe7ed848727ad8d8d04a9dd4332933cb9fad149d95}{69dc9d33c39b654dc20585fe7ed848727ad8d8d04a9dd4332933cb9fad149d95}\\
\href{https://mempool.space/tx/9508f0f37aa91abe92b66fc42a43ca801a7b4eca5d246fb7dd73db8eb91ad04e}{9508f0f37aa91abe92b66fc42a43ca801a7b4eca5d246fb7dd73db8eb91ad04e}\\
\href{https://mempool.space/fr/tx/af7e82e4ec88bbda055a4ee9319290635e7cac1a4d45f131d1182a5922a7d22b}{af7e82e4ec88bbda055a4ee9319290635e7cac1a4d45f131d1182a5922a7d22b}\\
\href{https://mempool.space/fr/tx/46e876485226b620e11a4d64bbbff92d9f0fd63161d8e68de0c13252f8187dc8}{46e876485226b620e11a4d64bbbff92d9f0fd63161d8e68de0c13252f8187dc8}\\
\href{https://mempool.space/fr/tx/f7dee8685741f68e33d28c4de0d4f43cfbf4dbb24794a4eb392cf456344fef1c}{f7dee8685741f68e33d28c4de0d4f43cfbf4dbb24794a4eb392cf456344fef1c}\\
\href{https://mempool.space/fr/tx/ae70f72a454ef7016c828db6d33a92550c797958f73856a220eae23b93da0588}{ae70f72a454ef7016c828db6d33a92550c797958f73856a220eae23b93da0588}\\
\href{https://mempool.space/fr/tx/9b9fd0150a1249c8c4e403b41f47070f52e1d0538bfc422b1684febe50cda103}{9b9fd0150a1249c8c4e403b41f47070f52e1d0538bfc422b1684febe50cda103}\\
\href{https://mempool.space/fr/tx/b5853bb7ce02e915aa8eac59b1256f1599247b27872b0b0786550c580dead652}{b5853bb7ce02e915aa8eac59b1256f1599247b27872b0b0786550c580dead652}\\
\href{https://mempool.space/fr/tx/4a13a422d8da493664be515fd8175775b77fe04aee599695275e5ecd6fae720f}{4a13a422d8da493664be515fd8175775b77fe04aee599695275e5ecd6fae720f}\\
\href{https://mempool.space/fr/tx/7dcd0538ad7bdcdb79d7d0244cf4ff002b46779e5594fe8b84fd1939d9770999}{7dcd0538ad7bdcdb79d7d0244cf4ff002b46779e5594fe8b84fd1939d9770999}\\
\href{https://mempool.space/fr/tx/c82f14c3e1bdf9651dc1f7521d8624586c3d4c603dd03dc43890531ba41da0ac}{c82f14c3e1bdf9651dc1f7521d8624586c3d4c603dd03dc43890531ba41da0ac}\\
\normalsize

\section{Referenced addresses}
\label{sec:appendix_referenced_addresses}
We present in this section the complete addresses of the addresses referenced in
the main text.
\\\\
\Small
\noindent
\href{https://mempool.space/address/bc1qx56r2dgqy8usgpg9qqqsqtqpqq9qq8sqqyqqqqs9sj86j6c9qqssptq452}{bc1qx56r2dgqy8usgpg9qqqsqtqpqq9qq8sqqyqqqqs9sj86j6c9qqssptq452}
\normalsize

\section{Dataset}
\label{sec:appendix_dataset}
\normalsize
We describe the datasets and artifacts that will be available upon request to the corresponding author.

\subsection{All Bitcoin addresses dataset}
\label{subsec:appendix_all_bitcoin_addresses}
This dataset corresponds to the 1,283,997,050 Bitcoin addresses that we
extracted from the blockchain up to block 840,682. This file is enhanced with
statistics on the addresses that we computed by processing all
transactions on the blockchain, we show them in columns.  It is a 2GB parquet
file containing a pandas DataFrame with 1,283,997,050 rows and 6 columns.
\begin{itemize}
    \item \texttt{address}: the Bitcoin address.
    \item \texttt{first apparition}: the UTC timestamp of the block in which this address appeared first.
    \item \texttt{last apparition}: the UTC timestamp of the block in which this address appeared Last.
    \item \texttt{number of transactions}: the number of transactions in which this address appeared.
    \item \texttt{total received}: the total amount of bitcoins received by this address in satoshis.
    \item \texttt{total sent}: the total amount of bitcoins sent by this address in satoshis.
\end{itemize}

\subsection{Ground truth initial dataset}
\label{subsec:appendix_ground_truth_dataset}
This ground truth dataset corresponds to the 208,656 bitcoin addresses with a
Shannon entropy strictly lower than 4 that we manually classified. This file is
enhanced with statistics on the addresses that we computed by processing all
transactions on the blockchain, we show them in columns. It is a CSV file with
208,656 rows and 8 columns.
\begin{itemize}
    \item \texttt{address}: the Bitcoin address.
    \item \texttt{manual classification}: the label of the manual classification of the address (burn, regular, or other).
    \item \texttt{entropy}: the Shannon entropy of the address.
    \item \texttt{first apparition}: the UTC timestamp of the block in which this address appeared first.
    \item \texttt{last apparition}: the UTC timestamp of the block in which this address appeared Last.
    \item \texttt{number of transactions}: the number of transactions in which this address appeared.
    \item \texttt{total received}: the total amount of bitcoins received by this address in satoshis.
    \item \texttt{total sent}: the total amount of bitcoins sent by this address in satoshis.
\end{itemize}

\subsection{Ground truth burn addresses}
\label{subsec:appendix_ground_truth_burn_addresses}
This dataset corresponds to the 9,672 burn addresses that we manually classified
in the ground truth initial dataset and during all predictions described in Section~\ref{sec:results_model_predictions}. This
file is a CSV file with 9,672 rows and 2 columns.
\begin{itemize}
    \item \texttt{address}: the Bitcoin address.
    \item \texttt{encoding}: the encoding of the address (Base58 or Bech32).
\end{itemize}

\subsection{Trained model}
\label{subsec:appendix_trained_model}
The final trained model after the second reinforcement is a 9MB Python pickle
file. It is a MLPClassifier from the scikit-learn library in version 1.5.1. It
can be directly used to classify regular and burn addresses on a pandas Series
of addresses.

\subsection{Code to train and use the model}
\label{subsec:appendix_code}
We make available the Python code we used to train the model, and classify
addresses using the trained model. This code includes the address feature encoding using
optimized parallelizable Numba code.

\end{document}